\documentclass[sigconf]{acmart}
\usepackage{subfigure}

\AtBeginDocument{%
  }

\copyrightyear{2026}
\acmYear{2026}
\setcopyright{cc}
\setcctype{by-nc-nd}
\acmConference[CHI EA '26]{Extended Abstracts of the 2026 CHI Conference on Human Factors in Computing Systems}{April 13--17, 2026}{Barcelona, Spain}
\acmBooktitle{Extended Abstracts of the 2026 CHI Conference on Human Factors in Computing Systems (CHI EA '26), April 13--17, 2026, Barcelona, Spain}
\acmDOI{10.1145/3772363.3798740}
\acmISBN{979-8-4007-2281-3/2026/04}

\begin{document}

\title{Investigating Effective Uncertainty Visualizations for Ordinal Crowdsourced Data of Crowding Conditions}
\author{Bea Alexis Arcega}
\orcid{0009-0007-3047-5970}
\affiliation{%
  \institution{De La Salle University}
  \city{Manila}
  \country{Philippines}
}
\email{bea_arcega@dlsu.edu.ph}

\author{Annika Dominique Campos}
\orcid{0009-0003-0949-561X}
\affiliation{%
  \institution{De La Salle University}
  \city{Manila}
  \country{Philippines}
}
\email{annika_campos@dlsu.edu.ph}

\author{Kathleen Therese R. Cruz}
\orcid{0009-0006-4138-572X}
\affiliation{%
  \institution{De La Salle University}
  \city{Manila}
  \country{Philippines}
}
\email{kathleen_cruz@dlsu.edu.ph}

\author{Aaron Ace R. Toledo}
\orcid{0009-0005-4266-0431}
\affiliation{%
  \institution{De La Salle University}
  \city{Manila}
  \country{Philippines}}
\email{aaron_toledo@dlsu.edu.ph}

\author{Briane Paul V. Samson}
\orcid{0000-0002-0253-452X}
\affiliation{%
  \institution{De La Salle University}
  \city{Manila}
  \country{Philippines}}
\email{briane.samson@dlsu.edu.ph}
\renewcommand{\shortauthors}{Arcega et al.}

\begin{abstract}
Commuters often encounter crowding in railway systems, particularly in queues where passenger density varies throughout the day. This introduces uncertainty in crowdedness, making it difficult for individuals to anticipate conditions and plan their trips effectively. Crowdsourcing has been a valuable method for collecting localized user data. But the unpredictability of crowds and the uncertainty of crowdsourced information pose new challenges for decision-making. However, we know little about how to effectively visualize uncertainty in crowdedness to support informed commuting decisions, particularly when using crowdsourced ordinal data. Here, we investigated different uncertainty visualizations and their effectiveness in representing the variability and reliability of crowdsourced crowding data. They were evaluated through an online study, and we found that cluster visualization is best suited to reduce cognitive load while maximizing user confidence and trust. On the other hand, participants showed higher accuracy in determining crowd levels when using bubble treemaps. 
\end{abstract}






\maketitle

\section{Introduction}
Uncertainty is a common aspect of daily life that affects how people think, feel, and respond to various situations, often arising from unpredictable events or insufficient information \cite{Walker2013}. As uncertainty increases, individuals often feel less confident in anticipating change, which can lead to anxiety, hesitation, and poor or misguided decisions \cite{marchau2019decision}. 


Uncertainty in public transit can be broadly defined as a lack of deterministic knowledge, whether from unpredictable schedules or crowding \cite{walker2003}. Crowding, particularly in urban rail systems, remains a common issue in high-density cities, especially during peak hours \cite{cheng2020analysis}. While peak-time crowding is expected, congestion during non-peak hours adds another layer of unpredictability, making it more difficult for commuters to anticipate conditions and plan their trips effectively. These inefficiencies, often tied to inadequate transit systems, lead to long commutes and passenger frustration \cite{chowdhury2015effects}. Beyond physical discomfort, crowding also causes psychological and emotional distress, including anxiety, fatigue, privacy concerns, health concerns, and fear for personal safety \cite{tirachini2017estimation, cox2006rail, cheng2010exploring, mahudin2011modelling, mahudin2012measuring, wardman2011twenty}.

Responses to crowding also vary, ranging from a simple changing of routes to a more complex decision of canceling travel altogether \cite{gentile2016modelling, tirachini2013crowding}. Social influence plays a role in how people respond to uncertainty, as observing the behavior of other commuters can shape individual decisions \cite{durand2018}. This may suggest that if many commuters choose to avoid a particular platform or visibly express discomfort, others are more likely to do the same, thus reinforcing a collective response to perceived discomfort or risk. The study addresses a gap in literature that has largely focused on quantitative crowd estimates in Western or high-income settings, without examining the unique behavioral, cultural, and infrastructural factors present in developing contexts. To mitigate the unpredictability of transit crowding, applications increasingly rely on crowdsourced data. In public transport, crowdsourcing is utilized in applications such as Waze, Moovit, and Google Maps, using user data to predict traffic and crowd levels \cite{apanasevic2021crowdsourcing}. Crowdsourcing is the process of outsourcing tasks from the collective ideas of various contributors through open calls \cite{zhen2021crowdsourcing} that could provide innovative solutions for various problems \cite{guazzini_modeling_2015}. 




While existing applications provide helpful data on current crowd levels, they do not account for uncertainty information. This means that users may still face unexpected conditions, as the application or platforms only provide point estimates; thus, do not show how reliable or complete the crowding data is. \cite{manski2019communicating} emphasized that presenting data without margins of error or confidence intervals can create an illusion of precision, leading users and policymakers to treat uncertain estimates as definitive truths. Such practices can ultimately result in suboptimal decisions, particularly when forecasts rest on assumptions that are not well supported by evidence. 

To address this gap, uncertainty visualizations can be utilized to convey uncertain information through visual cues \cite{zhao2023evaluating}. According to \cite{2018-uncertainty-bus}, incorporating uncertainty displays into public transportation systems can lead to more informed and higher-quality commuter decisions. They also emphasized that for these visualizations to be effective, the representations must be designed in a way that is easily understandable by non-expert users. Despite their potential, existing papers revealed gaps in understanding how visualizations contribute to human judgment and decision-making. Moreover, professionals across multiple disciplines continue to struggle with reasoning when presented with uncertainty visualizations \cite{padilla2021uncertain}, thus raising the question of whether commuters face similar difficulties when relying on these visual tools. 

With this in mind, this study identified which uncertainty visualizations are most effective at communicating uncertainty in crowdsourced information presented using ordinal categories of crowding. This paper specifically evaluated Visual Entropy, Bubble Treemaps, Icon Arrays, Hypothetical Outcome Plots (HOPs), and Clustering. To effectively evaluate these visualizations in the context of commuter decision-making, this study adopted an online survey grounded in five dimensions proposed by \cite{brennentuerk2018} and \cite{hullman2019}: accuracy, reported confidence, perceived trust, response time, and cognitive load. Our findings revealed that the Bubble Treemap and Clustering visualizations provided the most effective balance of accuracy and usability for interpreting ordinal uncertainty data. The findings are expected to inform future visualization techniques and commuter information systems that embrace uncertainty as a key design consideration. Ultimately, the study seeks to promote more transparent and human-centered approaches to information design, empowering commuters to navigate the inherent uncertainty of public transit with greater clarity and confidence.

\section{Related Works} 
Uncertainty is a daily phenomenon that spans from weather patterns to transit systems. In public transit, it is defined as a lack of deterministic knowledge about crowding or schedules \cite{walker2003}. On a psychological aspect, crowding uncertainty causes stress, anxiety, and discomfort that result in a change of routes, postponement of trips, and the development of long-term coping strategies \cite{kumagai2021commuters}. However, relying only on intuition is not sufficient for systematic decision-making. As uncertainty remains inherent in data, implementing visualization approaches that represent probabilistic phenomena is important to avoid errors in judgment and decision-making behavior \cite{spiegelhalter2011visualizing, padilla2021uncertain}. As stated by \cite{padilla2021uncertain}, a design space that balances statistical accuracy with cognitive accessibility is used to address uncertainty in the context of public transit.

There are numerous cognitive theories that shape the design space in the area of uncertainty visualizations. One approach is frequency framing, which suggests that people can easily understand uncertainty when presented as natural frequency than probability. These frequency formats are useful particularly for people with low numeracy skills \cite{padilla2021uncertain}. Conversely, in attribute substitution, people frequently interpret static visualizations, such as error bars and box plots, as absolute values instead of probabilistic ranges that hide the underlying features of the data \cite{brodlie2012review}. Additionally, designers need to be cautious of visual boundaries similar to the cone of uncertainty shown in weather forecasts, as it creates rigid cognitive categories, thereby dismissing areas of risk outside a boundary \cite{padilla_decision_2018}. 

Based on key theories discussed by \cite{padilla2021uncertain}, there are multiple ways to visualize uncertainty. First, icon arrays apply frequency framing by showing uncertainty as “X out of Y,” allowing users to interpret information through simple visual counting. Hypothetical outcome plots (HOPs) address attribute substitution by using animation to display variation over time. Similarly, \cite{holliman2019} proposed visual entropy, which uses 2D and 3D shapes in creating an ordinal scale designed for non-technical users. It applies visual cues like shape to represent uncertainty in a way that aligns with intuitive perception. \cite{gortler2018} introduced the bubble treemap, which displays uncertainty in hierarchical data using circular diagrams. Bubble treemaps also use visual semiotics, relying on size and contour to communicate uncertainty and data relationships. \cite{newen2022} discussed the use of clustering for identifying global relationships in high-dimensional data. It also incorporates spatial arrangement and avoids the issue of creating cognitive categories by presenting uncertainty without visual boundaries. In addition, visual metaphors such as fuzziness, transparency, spatial arrangement, and value-suppressing color palettes are part of a theoretical approach discussed by \cite{padilla2021uncertain}, which suggests that uncertainty visualizations are more effective when they align with how people intuitively interpret data. 

While these visualizations provide intuitive metaphors for uncertainty, there is no universal method that guarantees optimal judgment \cite{padilla2021uncertain}. Thus, choosing the most appropriate visualization method is based on the audience, decision-making context, and the nature of the uncertainty being conveyed. According to \cite{brennentuerk2018}, evaluating user understanding and reactions to visual uncertainty can be effectively guided by a structured, empirical framework. This framework examines four key factors: (1) how accurately users interpret the information, (2) the confidence they report in their interpretations, (3) the time they take to respond, and (4) the mental effort involved. Together, these dimensions offer a comprehensive view of the cognitive processes in interpreting uncertainty in visual representation. Aligned with this perspective,\cite{hullman2019} also examined additional factors, including perceived trust, perceived usefulness, and the impact of the visualizations on user behavior. 

To effectively evaluate these visualizations in the context of commuter decision-making, this study adopts a comprehensive, user-centered evaluation approach. These constructs will provide insight into both cognitive performance and subjective experience, ensuring that the chosen visualizations are not only informative but also intuitive, trustworthy, and supportive of real-world decision-making under uncertainty.

\section{Method}


This study utilized a quantitative data collection using an online survey to assess commuter understanding of different uncertainty visualizations and visual cues. 

\subsection{Participants}
We distributed the online survey to closed networks, online groups, and social media platforms, aimed at students and professionals aged 18 and older, particularly regular train commuters in a Philippine metropolitan area. We received 163 responses and efforts were made to include a diverse set of commuters based on age, commuting frequency, and occupation to ensure a range of perspectives. We removed responses that were potentially providing low-quality data due to straight-lining across all Likert questions and having an unusually short/fast average response time.  

\subsection{Materials and Protocols}



A design space was constructed through an extensive review of empirical studies and theoretical frameworks in uncertainty visualization. A total of five visualization types were selected to represent a diverse set of mechanisms for encoding uncertainty: frequency framing (Icon Arrays), spatial grouping (Clustering), visual semiotics and size (Bubble Treemaps), visual blur (Entropy), and temporal animation (HOPs).

\begin{figure*}[t]
   \centering                     
   \includegraphics[width=\textwidth]{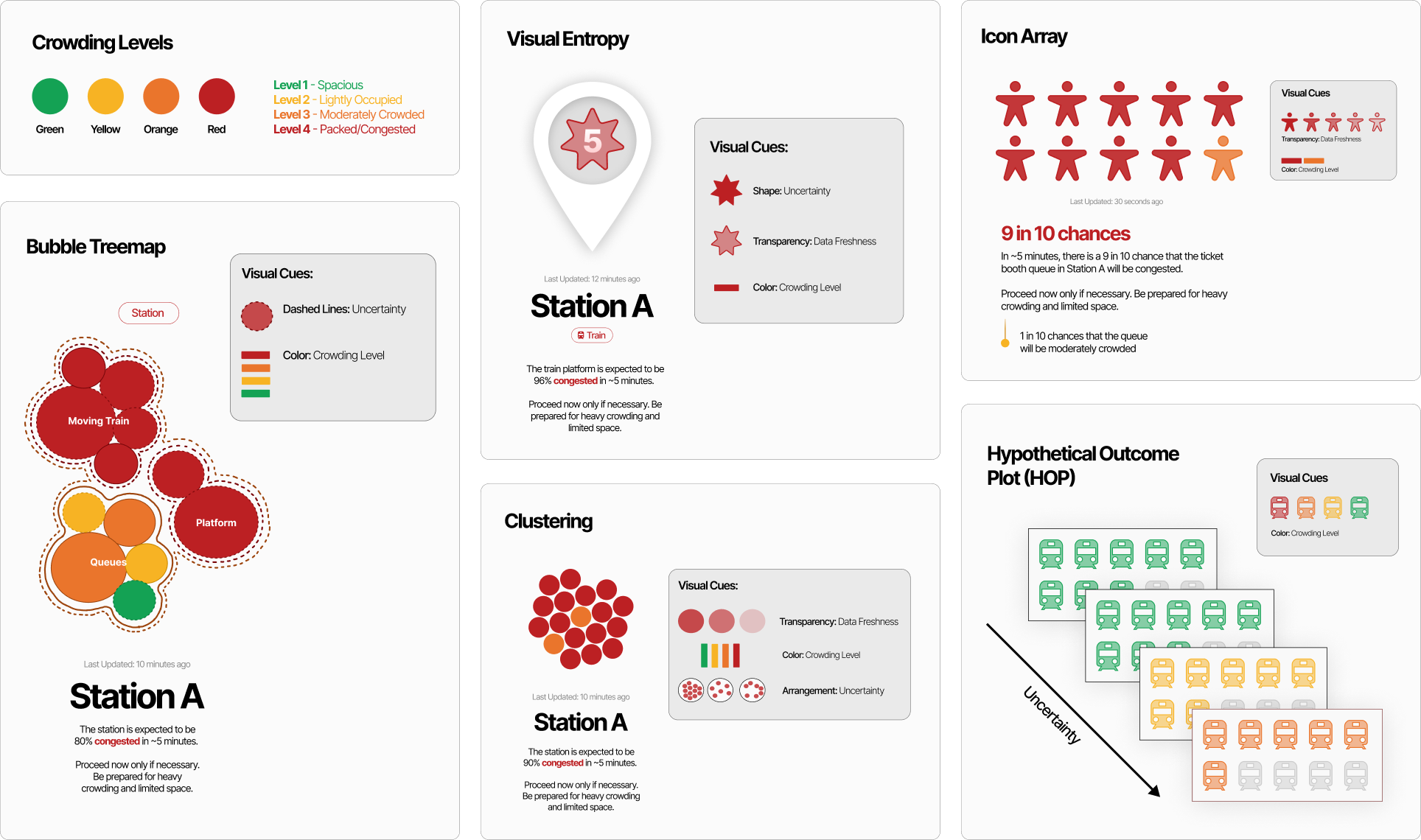}
   \caption{Our design space of uncertainty visualizations for crowding level reports. We primarily encoded crowding levels using four colors and explored the use of bubble treemaps, visual entropy, clustering, icon arrays, and hypothetical outcome plots.}
   \Description{This figure shows the we encoded crowding levels using four colors (green, yellow, orange, and red) and explored the use of bubble treemaps, visual entropy, clustering, icon arrays, and hypothetical outcome plots.}
   \label{fig:design-space}
\end{figure*}

To evaluate these techniques, crowding conditions were operationalized into a four-level ordinal scale: Spacious, Lightly Occupied, Moderately Congested, and Congested. This specific scale was chosen because it reflects intuitive, real-world passenger density levels that everyday commuters can easily perceive, avoiding the cognitive burden of exact numerical estimates.
Furthermore, to ensure rigorous experimental control, the amount and granularity of information were held constant across all conditions. For any given task, all five visualizations represented the exact same underlying probabilistic distribution of data across the four categories. Therefore, observed differences in user performance are strictly attributable to the visual encoding strategies rather than variations in information quantity. Figure \ref{fig:design-space} illustrates the selected uncertainty visualizations in representing different crowding conditions. These visualizations were chosen to represent a diverse set of mechanisms for encoding uncertainty in the context of crowdsourced, categorical and ordinal data. 

To assess interpretability of these visualizations, an online survey was conducted through Jotforms, which gathered commuter responses about crowding perception and their understanding of uncertainty visualizations and visual cues. The online survey begins with obtaining participant consent and providing a brief overview of the study’s purpose. Participants answered a demographic and commuting behavior survey, followed by a sequence of five interpretation tasks. For each visualization, participants were asked to determine the perceived level of crowding uncertainty using a multiple-choice scale, followed by a set of Likert-scale items that assess their confidence, trust, and cognitive effort. Response time was automatically recorded for each task to provide an additional measure of interpretability. The goal of the survey was to evaluate accuracy, reported confidence, response time, cognitive load, and trust.



\subsection{Measures}

For the online study, the five metrics from \cite{brennentuerk2018} and \cite{hullman2019} will be used to assess the interpretation of the participants and the general understanding of uncertainty visualizations.

To provide a clear method for determining which visualization approach is effective, a systematic quantitative analysis will be performed on the results. The objective of this phase is to identify every visualization technique from our design space that proves to be effective for implementation. This process begins with calculating a mean score  for accuracy and median for response time, reported confidence, cognitive load, and trust.

Participant interpretation was evaluated across five key metrics. Accuracy was assessed based on the participant's ability to correctly identify the ground-truth crowding or uncertainty level. A binary scoring logic was applied, where a response was marked correct only if it matched the intended mode of the crowding distribution or the predetermined uncertainty level on the five-point scale. Cognitive load, confidence, and trust were assessed using 5-point Likert scales ranging from strongly disagree to strongly agree. These items captured the ease of understanding, self-reported certainty, and perceived reliability of each visualization. Finally, response time was recorded as the duration in seconds taken to complete each task.

To determine the most effective visualization, an Effectiveness Score was formulated. Accuracy was prioritized as the primary metric with a 40\% weight, followed by Cognitive Load and Response Time at 20\% each, and Confidence and Trust at 10\% each. This weighting reflects the critical need for commuters to reach correct interpretations while maintaining a reasonable mental workload. Sensitivity analysis confirmed that Bubble Treemap and Cluster remained the top two visualizations regardless of whether weights were shifted towards higher accuracy or lower cognitive load, demonstrating the stability of our findings. This selection will be validated using statistical tests through Kruskal-Wallis H-test to confirm the performance differences are statistically significant.

\section{Results}

\begin{figure}[h]
\centering
\includegraphics[width=1\linewidth]{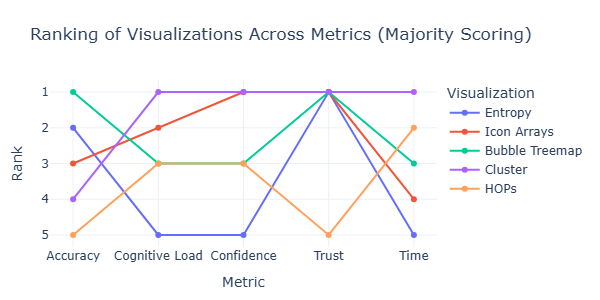}
\caption{The ranking of uncertainty visualizations based on accuracy, cognitive load, confidence, trust and response time (majority scoring).}
\Description{The figure shows the rank of visualizations across metrics using majority scoring. This ranking was based on accuracy, cognitive load, confidence, trust, and response time.}
\label{fig:ranking-plot}
\end{figure}

To provide an overview of visualization performance, we calculated the effectiveness score for each technique. This score represents a weighted synthesis of objective performance (Accuracy at 40\%, Response Time at 20\%) and subjective user experience (Cognitive Load at 20\%, Confidence at 10\%, and Trust at 10\%). Points were assigned based on relative ranking (5 points for 1st place, down to 1 point for 5th). Based on this evaluation, the Bubble Treemap emerged as the most effective overall score (4.0), primarily due to its dominance in accuracy. It was followed by Cluster (3.8), Icon Array (3.4), Entropy (2.6), and HOPS (2.2). The relative rankings across individual metrics are visualized in Figure~\ref{fig:ranking-plot}. Table~\ref{tab:summary-results} provides the descriptive and inferential statistics for these metrics.

\begin{table}[h]
    \centering
    \caption{Summary of Results: Descriptive Statistics and Kruskal–Wallis Test ($df=4$)}
    \label{tab:summary-results}
    \small 
    \begin{tabular}{@{}lccccc@{}}
        \toprule
        \textbf{Visualization} & \textbf{Acc. (\%)} & \textbf{Cog. (M)} & \textbf{Conf. (M)} & \textbf{Trust (M)} & \textbf{Time (s)} \\
        \midrule
        Entropy         & 64.00          & 2.50          & 3.00          & 4.00          & 233 \\
        Icon Arrays     & 53.01          & 3.50          & 4.00          & 4.00          & 180 \\
        Bubble Treemap  & \textbf{69.28} & 3.00          & 3.50          & 4.00          & 87  \\
        Cluster         & 51.81          & \textbf{4.00} & \textbf{4.00} & \textbf{4.00} & \textbf{46}  \\
        HOPS            & 32.53          & 3.00          & 3.50          & 3.00          & 48  \\
        \midrule
        \textbf{K-W ($H$)} & 89.48 & 70.32 & 38.92 & 21.82 & 424.15 \\
        \textbf{Effect ($\eta^2$)} & 0.11 & 0.08 & 0.04 & 0.02 & 0.52 \\
        \bottomrule
        \specialrule{0pt}{2pt}{0pt}
        \multicolumn{6}{l}{\footnotesize \textit{Note:} Acc. is Mean \%; others are Median (M) values. All $p < 0.001$.}
    \end{tabular}
\end{table}

To test for significant differences in performance, we conducted a Kruskal-Wallis $H$-test for each metric. As shown in Table~\ref{tab:summary-results}, results confirm statistically significant differences across all evaluated metrics. Notably, the choice of visualization exerted a massive effect on Response Time ($\eta^2=0.52$) and a moderate-to-large effect on Accuracy ($\eta^2=0.11$).

To identify specific performance differences, we performed Dunn’s post-hoc pairwise comparisons (summarized in Table~\ref{tab:full-posthoc}). This analysis reveals a clear trade-off between objective performance and subjective user experience.

The \textit{Bubble Treemap} was the most accurate technique (mean = 69.28\%), significantly outperforming Icon Arrays (53.01\%, $r=-0.23$) and HOPS (32.53\%, $r=0.47$). While the \textit{Cluster} visualization enabled the fastest task completions (medians of 46s and 48s, respectively), it was significantly less accurate than the Bubble Treemap ($r=0.14$). These accuracy gains in the Treemap came at the cost of speed; \textit{Entropy} (233s) and \textit{Icon Arrays} (180s) were significantly slower than all other groups ($p < 0.001$).

\begin{table*}[t]
  \centering
  \small
  \caption{Post-Hoc Pairwise Comparisons (Dunn’s Test)}
  \label{tab:full-posthoc}
  \begin{tabular}{llcc p{0.6cm} llcc} 
    \toprule
    \textbf{Metric} & \textbf{Comparison} & \textbf{\textit{p}} & \textbf{\textit{r}} & & \textbf{Metric} & \textbf{Comparison} & \textbf{\textit{p}} & \textbf{\textit{r}} \\
    \midrule
    \textbf{Accuracy} & Entropy vs. Icon & 0.041 & 0.19 & & \textbf{Cog. Load} & Entropy vs. Icon & $<$0.001 & -0.28 \\
     & Entropy vs. HOPS & $<$0.001 & 0.45 & & & Entropy vs. Cluster & $<$0.001 & -0.42 \\
     & Icon vs. Bubble & 0.002 & -0.23 & & & Entropy vs. HOPS & 0.001 & -0.22 \\
     & Icon vs. HOPS & $<$0.001 & 0.33 & & & Icon vs. Cluster & 0.021 & -0.18 \\
     & Bubble vs. Cluster & 0.011 & 0.14 & & & Bubble vs. Cluster & $<$0.001 & -0.31 \\
     & Bubble vs. HOPS & $<$0.001 & 0.47 & & & Cluster vs. HOPS & 0.001 & 0.21 \\
     \cmidrule{6-9}
     & Cluster vs. HOPS & 0.001 & 0.18 & & \textbf{Confidence} & Entropy vs. Icon & 0.004 & -0.20 \\
    \cmidrule{1-4} 
    \textbf{Resp. Time} & Entropy vs. Bubble & $<$0.001 & 0.55 & & & Entropy vs. Cluster & $<$0.001 & -0.31 \\
     & Entropy vs. Cluster & $<$0.001 & 0.76 & & & Bubble vs. Cluster & $<$0.001 & -0.24 \\
     & Entropy vs. HOPS & $<$0.001 & 0.79 & & & Cluster vs. HOPS & 0.001 & 0.21 \\
     \cmidrule{6-9}
     & Icon vs. Bubble & $<$0.001 & 0.48 & & \textbf{Trust} & Entropy vs. Cluster & 0.011 & -0.17 \\
     & Icon vs. Cluster & $<$0.001 & 0.73 & & & Bubble vs. Cluster & 0.002 & -0.20 \\
     & Icon vs. HOPS & $<$0.001 & 0.77 & & & Cluster vs. HOPS & 0.001 & 0.21 \\
     & Bubble vs. Cluster & $<$0.001 & 0.43 & & & & & \\
     & Bubble vs. HOPS & $<$0.001 & 0.46 & & & & & \\
    \bottomrule
    \specialrule{0pt}{2pt}{0pt}
    \multicolumn{9}{l}{\footnotesize \textit{Note:} Only significant pairs ($p < 0.05$) are shown. Effect sizes ($r$) are based on the Wilcoxon Rank-Sum $Z$ statistic.}
  \end{tabular}
\end{table*}

The \textit{Cluster} visualization emerged as the superior choice for subjective user experience. It imposed the lowest cognitive load ($\text{median} = 4.0$), significantly outperforming Entropy ($r=-0.42$) and the Bubble Treemap ($r=-0.31$). Furthermore, while Figure~\ref{fig:ranking-plot} shows a tie in median Trust scores (4.00) across four techniques, the post-hoc analysis revealed that the distribution of responses for Clustering was significantly higher than the Bubble Treemap ($r=-0.20$) and Entropy ($r=-0.17$). This suggests that the spatial grouping of the Cluster visualization effectively lowers mental effort and fosters greater user assurance, even when it is less precise for raw data extraction.

\section{Discussion}
The findings reveal a significant perception-performance gap such that the design producing the highest objective accuracy (Bubble Treemap) is not the one that fosters the greatest user trust (Cluster). The superior accuracy of the Bubble Treemap suggests that its reliance on visual semiotics (size and contour) provides a highly effective framework for data extraction \cite{gortler2018}. By presenting uncertainty through a static arrangement, it allows users to use frequency framing without the cognitive burden of temporal integration \cite{padilla2021uncertain}. 

However, a striking disparity emerges regarding user experience. Although participants were most accurate with the Bubble Treemap, they reported significantly higher levels of confidence and trust when using the Cluster visualization. Following \cite{padilla_decision_2018}, we suggest that rigid boundaries, such as the circular contours in a treemap, can create a sense of false precision that increases user anxiety. In contrast, the Cluster visualization utilizes spatial arrangement without hard boundaries, presenting uncertainty as a global relationship rather than discrete containers \cite{newen2022}. This representation appears to align more closely with intuitive metaphors for uncertainty, lowering cognitive load and fostering user assurance. In high-stress transit environments, commuters may prefer a representation that feels honest over a precise one that feels mentally taxing.

While HOPs address attribute substitution through animation, they resulted in the lowest accuracy (32.53\%) and were not significantly faster than Clustering. This confirms that temporal visualizations require significant cognitive effort to mentally average random draws \cite{hullman2019}, overwhelming working memory during split-second decision-making. Similarly, while Icon Arrays utilize accessible frequency framing, the serial nature of visual counting made them significantly slower (180s) than spatial techniques.

These results suggest there is no universal method for transit uncertainty, but rather a choice between precision and comfort. If the primary goal is mathematical accuracy, the Bubble Treemap is the ideal choice. Conversely, if the goal is to minimize commuter stress and build trust, the Cluster visualization is superior as it reduces mental effort and aligns with intuitive perception.

\section{Conclusion} 

The statistical analysis confirms that the choice of visualization significantly impacts a user’s ability to interpret ordinal crowdsourced uncertainty data. Our findings demonstrate a distinct trade-off between mathematical precision and user comfort, suggesting that there is no universal "best" visualization for all commuting scenarios. While the Bubble Treemap emerged as the most effective overall (4.0) due to its superior objective accuracy (69.28\%), it was slightly slower and less trusted than the Cluster visualization. Conversely, the Cluster visualization (3.8) proved to be the most human-centered design for high-stress decision-making, as it achieved the fastest task completions (median = 46s) and significantly reduced cognitive load while maximizing user trust. 

While Icon Arrays utilized frequency framing to make information accessible, the serial nature of visual counting resulted in significantly slower response times (180s), making them less efficient for real-time applications. Similarly, both HOPs and Entropy generally performed poorly in this context, with HOPs yielding the lowest accuracy (32.53\%) and Entropy requiring the longest interpretation time (233s). We conclude that for systems where mathematical accuracy is the primary goal, the semiotic strength of the Bubble Treemap makes it the ideal choice. However, for real-time station displays where speed and user assurance are prioritized, the spatial grouping of the Cluster visualization is superior. These insights provide a guide for transit developers to move beyond point estimates and design more transparent, trust-based information systems that consider the specific cognitive constraints of the modern commuter.
 



\bibliographystyle{ACM-Reference-Format}
\bibliography{sample-base.bib}

\end{document}